\documentclass[final]{svjour2}
\usepackage{graphicx}
\usepackage{rotating}
\usepackage{amssymb}
\usepackage{mathptmx}
\usepackage[numbers]{natbib}
\usepackage{bm}
\makeatletter
\journalname{Journal of Low Temperature Physics}

\def\dag{\dagger}
\def\up{\uparrow}
\def\dn{\downarrow}
\def\sig{\sigma}
\def\Sig{\Sigma}
\def\om{\omega}
\def\Gam{\Gamma}

\def\lesssim{\ \raise.3ex\hbox{$<$}\kern-0.8em\lower.7ex\hbox{$\sim$}\ }
\def\gesim{\ \raise.3ex\hbox{$>$}\kern-0.8em\lower.7ex\hbox{$\sim$}\ }

\bibpunct{}{}{,}{s}{}{,}

\begin{document}

\newcommand{\hdblarrow}{H\makebox[0.9ex][l]{$\downdownarrows$}-}
\title{Superfluid phase transition and strong-coupling effects in an ultracold Fermi gas with mass imbalance}

\author{R. Hanai \and T. Kashimura \and R. Watanabe \and D. Inotani \and Y. Ohashi}
\institute{Department of Physics, Faculty of Science and Technology, Keio University,\\ 
3-14-1, Hiyoshi, Kohoku-ku, Yokohama 223-8522, Japan\\
\email{rhanai09@gmail.com}
}

\date{7.7.2012}

\maketitle



\begin{abstract}

We investigate the superfluid phase transition and effects of mass imbalance in the BCS (Bardeen-Cooper-Schrieffer)-BEC (Bose-Einstein condensation) crossover regime of an cold Fermi gas. We point out that the Gaussian fluctuation theory developed by Nozi\`eres and Schmitt-Rink and the $T$-matrix theory, that are now widely used to study strong-coupling physics of cold Fermi gases, give unphysical results in the presence of mass imbalance. To overcome this problem, we extend the $T$-matrix theory to include higher-order pairing fluctuations. Using this, we examine how the mass imbalance affects the superfluid phase transition. Since the mass imbalance is an important key in various Fermi superfluids, such as $^{40}$K-$^6$Li Fermi gas mixture, exciton condensate, and color superconductivity in a dense quark matter, our results would be useful for the study of these recently developing superfluid systems.

\end{abstract}
\section{Introduction}

The realization of a superfluid Fermi gas with mass imbalance is an exciting challenge in cold atom physics. Once this pairing state is achieved, using a tunable interaction associated with a Feshbach resonance, we can study various properties of this novel superfluid phase from the weak-coupling BCS regime to the strong-coupling BEC limit in a unified manner\cite{Giorgini}. Indeed, the so-called Sarma phase\cite{Sarma} discussed in a spin-polarized Fermi superfluid has also been predicted in this system\cite{Baarsma}. Since the formation of Cooper pairs between different species has been discussed in various systems, such as an exciton condensate in a semiconductor\cite{Yoshioka,Kasprzak}, and color superconductivity in a dense quark matter\cite{Barrois}, the realization of mass imbalanced superfluid Fermi gas would contribute to the further developments of these recently developing research fields. So far, the superfluid phase has not been realized in a mass-imbalanced Fermi gas. However, the formation of hetero-molecules was recently achieved in a $^{40}$K-$^6$Li Fermi-Fermi mixture\cite{Taglieber}, so that their superfluid phase transition might be realized near future. 
\par
In this paper, we investigate the superfluid phase transition of an ultracold Fermi gas with mass imbalance in the BCS-BEC crossover region. In the ordinary {\it mass-balanced} case, the strong-coupling theory developed by Nozi\`{e}res and Schmitt-Rink (NSR)\cite{Nozieres}, as well as the $T$-matrix approximation (TMA)\cite{Rohe, Perali, Tsuchiya}, have been extensively used to successfully explain various aspects of the BCS-BEC crossover physics. However, we show that these strong-coupling theories do not work well in the presence of mass imbalance. A similar problem has also been pointed out in a superfluid Fermi gas with population imbalance\cite{Parish,Liu}. In this paper, we extend the $T$-matrix theory to include higher order pairing fluctuations, so as to overcome this serious problem. Using this extended $T$-matrix theory, we examine how the mass imbalance affects the superfluid phase transition in the BCS-BEC crossover region.
\par
\section{Extended $T$-matrix for a Fermi gas with mass imbalance}
\par
We consider a uniform two-component Fermi gas, described by the Hamiltonian,
\begin{equation}
H = \sum_{\bm{p},\sigma}\xi_{\bm{p}\sig}c^{\dag}_{\bm p\sig}c_{\bm p\sig}
-U\sum_{\bm q}\sum_{\bm p, \bm{p}'}
c^{\dag}_{\bm p + \bm q/2\up}c^\dag_{-\bm p + \bm q/2\dn}c_{-\bm {p}' + \bm q /2\dn}c_{\bm{p}'+\bm q /2}.
\label{HAM}
\end{equation}
Here, $c_{\bm p\sig}$ is an annihilation operator of a Fermi atom with momentum ${\bm p}$ and pseudospin $\sig = \up,\dn$ describing two atomic hyperfine states. $\xi_{\bm p\sig}=p^2/(2m_\sig) - \mu_\sig$ is the kinetic energy of the $\sigma$-spin component, measured from the Fermi chemical potential $\mu_{\sig}$ (where $m_{\sig}$ is the mass of a $\sig$-spin atom). The pairing interaction $-U$ $(<0)$ is assumed to be tunable by adjusting the threshold energy of a Feshbach resonance (although we have ignored the detailed Feshbach mechanism in Eq. (\ref{HAM})). As usual, we measure the interaction strength in term of the inverse scattering length $a_s$, defined by
\begin{equation}
\frac{4\pi a_s}{m}=\frac{-U}{1-U\sum_{\bm p}\frac{m}{p^2}},
\end{equation}
where $m$ is twice the reduced mass ($2m^{-1}=m^{-1}_{\up}+m^{-1}_{\dn}$). In this scale, the weak-coupling BCS regime and the strong-coupling BEC regime are, respectively, characterized by $(k_{\rm F}a_s)^{-1}\lesssim -1$ and  $1 \gesim (k_{\rm F}a_s)^{-1}$ (where $k_{\rm F}=(3\pi N)^{1/3}$ is the Fermi momentum of a gas of $N$ Fermi atoms). The region between $-1 \lesssim (k_{\rm F}a_s)^{-1}\lesssim 1$ is called the crossover region. 
\par
\begin{figure}
\begin{center}
\includegraphics[%
  width=0.8\linewidth,
  keepaspectratio]{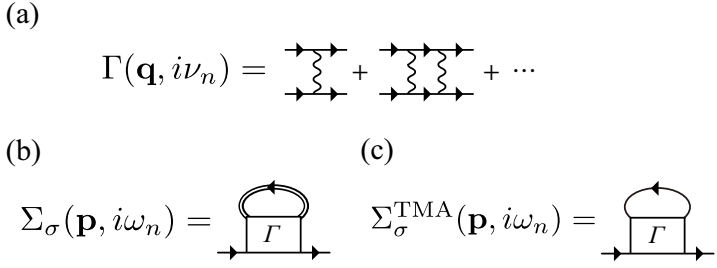}
\end{center}
\caption{(a) particle-particle scattering matrix $\Gamma({\bf q},i\nu_n)$ in the ladder approximation. (b) self-energy $\Sigma_\sig({\bm p}, i\om_n)$ in the extended $T$-matrix theory. (c) self-energy $\Sigma_\sig^{\rm{TMA}}({\bm p}, i\om_n)$ in the ordinary $T$-matrix theory. In this figure, the double solid line and the single solid line describe $G$ and $G^0$, respectively. The wavy line denotes the pairing interaction $-U$.}
\label{fig1}
\end{figure}
\par
To overcome the above mentioned breakdown of the NSR theory and $T$-matrix theory in the presence of mass imbalance, we extend the $T$-matrix theory to include higher order fluctuations in the Cooper channel. In our {\it extended} $T$-matrix approximation (ETMA), the self-energy $\Sigma_\sigma({\bf p},i\omega_n)$ in the single-particle Green's function,
\begin{equation}
G_{\bm p\sig}=\frac{1}{i\om_n - \xi _{\bm p\sig}-\Sig_\sig(\bm p,i\om_n)},
\label{etmag}
\end{equation}
has the form
\begin{equation}
\Sig_\sig(\bm p,i\om _n)=T\sum _{\bm q,\nu _n}\Gam(\bm q,i\nu _n)
G_{\bm q-\bm p,-\sig}(i\nu_n-i\om_n).
\label{etma}
\end{equation}
Here, $\omega_n$ and $\nu_n$ are the fermion and boson Matsubara frequencies, respectively. The particle-particle scattering matrix,
\begin{equation}
\Gam(\bm q, i\nu _n)=-{U  \over 1-U\Pi(\bm q,i\nu_n)},
\label{GAM}
\end{equation}
involves effects of pairing fluctuations within the ladder approximation, which is diagrammatically described by Fig.\ref{fig1}(a). Here, $G^0_{\bm p\sig}(i\om_n)=[i\om_n -\xi_{\bm p\sig}]^{-1}$ is the Green's function for a free Fermi gas, and
\begin{equation}
\Pi(\bm q, i\nu_n)=T\sum_{\bm p,i\om _n}G^0_{\bm p + \bm q/ 2,\up}(i\nu_n+i\om_n)G^0_{-\bm p+\bm q /2,\dn}(-i\om_n)
\label{PI}
\end{equation}
is the lowest-order pair-correlation function.
\par
In the ordinary (non-self-consistent) $T$-matrix approximation (TMA), the Green's function $G$ in Eq.(\ref{etma}) is replaced by the non-interacting Green's function as
\begin{equation}
\Sig_\sig^{\rm TMA}(\bm p,i\om _n)=
T\sum _{\bm q,\nu _n}\Gam(\bm q,i\nu _n)
G^0_{\bm q-\bm p,-\sig}(i\nu_n-i\om_n).
\label{tma}
\end{equation}
Equation (\ref{tma}) is also used in the NSR theory, but the Green's function in Eq. (\ref{etmag}) is expanded up to $O(\Sigma^{\rm TMA})$ as
\begin{equation}
G^{\rm{NSR}}_{\bm p\sig}(i\om_n)=G^0_{\bm p\sig}(i\om_n)
+ G^0_{\bm p\sig}(i\om_n) 
\Sig^{\rm TMA}_\sig(\bm p,i\om_n) G^0_{\bm p\sig}(i\om_n).
\label{NSRA}
\end{equation}
\par
The superfluid phase transition temperature $T_{\rm c}$ is conveniently determined from the Thouless criterion, which states that the phase transition occurs when the particle-particle scattering matrix $\Gamma({\bf q},i\nu_n)$ has a pole at ${\bf q}=\nu_n=0$. Since the three approximate theories, ETMA, TMA, and NSR, all use the same $\Gamma({\bf q},i\nu_n)$ in Eq. (\ref{GAM}), the resulting $T_{\rm c}$-equation is also the same among them, which is given by
\begin{equation}
1={U \over 2}\sum_{\bf p}
{\tanh(\xi_{{\bf p}\uparrow}/(2T))+\tanh(\xi_{{\bf p}\downarrow}/(2T)) 
\over \xi_{{\bf p}\uparrow}+\xi_{{\bf p}\downarrow}}.
\label{GAP}
\end{equation}
Since the chemical potential $\mu_\sigma$ is known to remarkably deviate from the Fermi energy in the BCS-BEC crossover, we actually solve the $T_{\rm c}$-equation (\ref{GAP}), together with the equations for the number $N_{\sigma={\uparrow,\downarrow}} (=N/2)$ of $\sigma$-spin atoms,
\begin{equation}
{N \over 2}=T\sum_{\bm p,i\om _n}G_{\bm p\sig}(i\om_n),
\end{equation}
and determine $T_{\rm c}$, $\mu_\uparrow$, $\mu_\downarrow$, self-consistently. 

\begin{figure}
\begin{center}
\includegraphics[%
  width=0.85\linewidth,
  keepaspectratio]{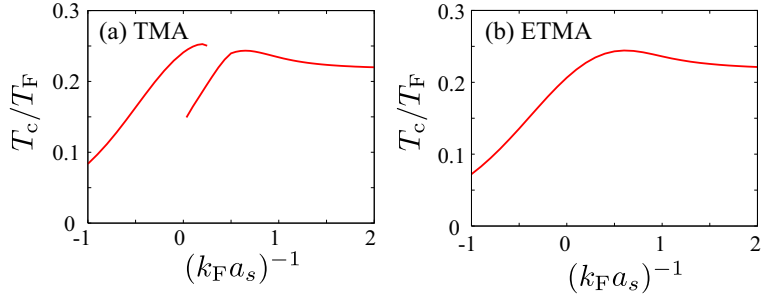}
\end{center}
\caption{(Color online) Calculated superfluid phase transition temperature $T_{\rm c}$ in the BCS-BEC crossover, normalized by $T_{\rm F}=(3\pi^2N/2\sqrt 2)^{2/3}m^{-1}$. (a) ordinary $T$-matrix approximation (TMA). (b) extended $T$-matrix approximation (ETMA). The ratio of the mass imbalance is taken to be $m_\uparrow/m_\downarrow=0.9$.}
\label{fig2}
\end{figure}

\section{Superfluid phase transition ans effects of mass imbalance}
\par
Figure \ref{fig2}(a) shows the superfluid phase transition temperature $T_{\rm c}$ for $m_\uparrow/m_\downarrow=0.9$, calculated within TMA (where the self-energy in Eq. (\ref{tma}) is used). Clearly, the calculated $T_{\rm c}$ exhibits unphysical behavior around the unitarity limit ($(k_{\rm F}a_s)^{-1}\sim 0$). A similar unphysical result for $T_{\rm c}$ is also obtained in the NSR theory (although we do not explicitly show the result here).
\par

\begin{figure}
\begin{center}
\includegraphics[%
  width=0.47\linewidth,
  keepaspectratio]{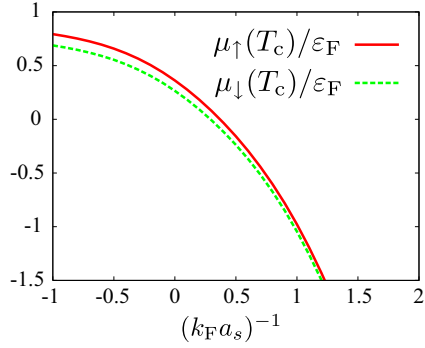}
\end{center}
\caption{(Color online) Calculated ETMA chemical potential $\mu_\sigma$ in the BCS-BEC crossover at $T_{\rm c}$, normalized by the Fermi energy $\varepsilon_{\rm F}=(3\pi^2N/2\sqrt 2)^{2/3}m^{-1}$ of a Fermi gas with mass $m$. We set $m_\up/m_\dn=0.9$.}
\label{fig3}
\end{figure}

To understand the origin of this unphysical result, it is helpful to compare the present system with a polarized Fermi gas, where the breakdown of the NSR theory and TMA have also been pointed out\cite{Liu,Parish,Kashimura}. For this purpose, we rewrite the kinetic energies $\xi_{{\bf p},\sigma=\uparrow,\downarrow}$ as,
\begin{eqnarray}
\xi_{\bm p\up}&=&\frac{m}{m_\up}\left(\frac{p^2}{2m}-\mu \right ) - h, \label{xiup}\\
\xi_{\bm p\dn}&=&\frac{m}{m_\dn}\left(\frac{p^2}{2m}-\mu \right ) + h, \label{xidn}
\end{eqnarray}
where $\mu=(\mu_\up + \mu_\dn)/2$, and 
\begin{equation}
h={m_\up\mu_\up-m_\dn\mu_\dn \over m_\up+m_\dn}.
\label{HHH}
\end{equation}
When $h=0$, the Fermi surfaces of the two components, determined by $\xi_{{\bf p},\sigma}=0$, coincide with each other. On the other hand, the mismatch of the Fermi surfaces occurs when $h\ne 0$. That is, $h$ effectively works as an external magnetic field applied to the system. In this sense, effects of mass imbalance is similar to those of population imbalance (where the mismatch of the Fermi surfaces between the two components is induced by setting $N_\uparrow\ne N_\downarrow$.) In the latter spin-polarized system, the reason for the breakdown of the NSR theory and TMA is known to be the incomplete treatments of magnetic fluctuations and pseudogap effects associated with pairing fluctuations\cite{Liu,Parish,Kashimura}. It has also been pointed out\cite{Kashimura} in the case of spin-polarized Fermi gas that one can eliminate this serious problem when the self-energy in Fig.\ref{fig1}(b) is used. Then, because of the similarity between the two systems, ETMA is expected to be also valid for the problem in the mass-imbalanced case. Indeed, as shown in Fig.\ref{fig2}(b), ETMA gives the expected smooth crossover behavior of $T_{\rm c}$ in the BCS-BEC crossover. That is, starting from the weak-coupling regime, $T_{\rm c}$ gradually increases with the increase of the interaction strength, and it approaches a constant value in the BEC regime. In the BEC limit, the system is well described by a gas of tightly bound hetero-molecules with the particle number $N/2$ and molecular mass $M=m_\uparrow+m_\downarrow$. Noting this, one finds 
\begin{equation}
T_{\rm c}={2m \over m_\uparrow+m_\downarrow}\times 0.218T_{\rm F}
~~~(\equiv T_{\rm BEC}),
\label{BEC}
\end{equation}
in the BEC limit (where $T_{\rm F}=(3\pi^2N/2\sqrt 2)^{2/3}m^{-1}$ is the Fermi temperature of a Fermi gas with the mass $m$).
\par
Figure \ref{fig3} shows the ETMA chemical potential $\mu_\sigma$ in the BCS-BEC crossover at $T_{\rm c}$. As expected from the ETMA result on $T_{\rm c}$ in Fig.\ref{fig2}(b), one sees smooth behavior of the chemical potential as a function of the interaction strength. As in the mass-balanced case, both  $\mu_\uparrow$ and $\mu_\downarrow$ gradually decrease, as one passes through the BCS-BEC crossover region. They become negative in the BEC regime, which is a typical phenomenon of the BCS-BEC crossover physics. 
\par

\begin{figure}
\begin{center}
\includegraphics[%
  width=0.95\linewidth,
  keepaspectratio]{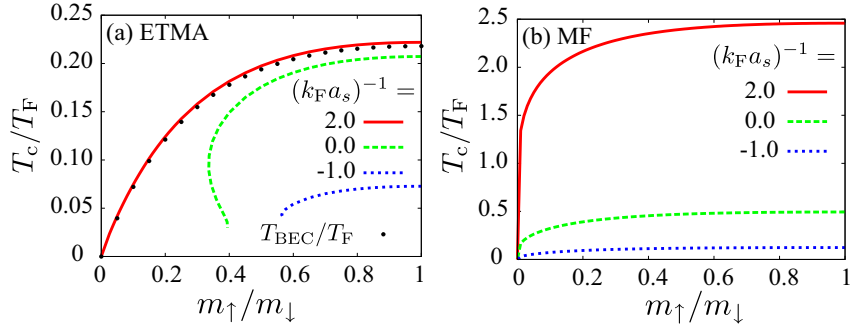}
\end{center}
\caption{(Color online) Superfluid phase transition temperature $T_{\rm c}$ as a function of the mass imbalance ratio $m_\up/m_\dn$. (a) extended $T$-matrix approximation (ETMA). (b) mean-field theory (MF). In this figure, $T_{\rm BEC}$ is the BEC phase transition temperature of an ideal hetero-molecular Bose gas, given by Eq. (\ref{BEC}).}
\label{fig4}
\end{figure}

\par
We now examine effects of the mass imbalance on the superfluid phase transition temperature $T_{\rm c}$ by using the extended $T$-matrix theory. As shown in Fig.\ref{fig4}, the mass imbalance tends to decrease $T_{\rm c}$, irrespective of the magnitude of the interaction strength. In the BEC regime ($(k_{\rm F}a_s)^{-1}=2$), a finite $T_{\rm c}$ is obtained down to the limit $m_\uparrow/m_\downarrow\rightarrow 0$. In this regime, hetero-molecules have already been formed above $T_{\rm c}$, the superfluid phase transition is dominated by the BEC of these molecules.  Indeed, as shown in Fig.\ref{fig4}(a), $T_{\rm c}$ in this regime is well described by the BEC phase transition temperature of an ideal Bose gas, given by $T_{\rm BEC}$ in Eq. (\ref{BEC}).
\par
In contrast to the BEC regime, $T_{\rm c}$ vanishes at a certain value of $m_\uparrow/m_\downarrow$ ($<1$) in the BCS regime, as shown in Fig. \ref{fig4}(a). At a glance, this vanishing $T_{\rm c}$ seems to be because of the depairing effect by the effective magnetic field $h$ in Eq. (\ref{HHH}). However, when we substitute the Fermi energy $\varepsilon_{{\rm F}\sigma}=(3\pi^2N/2\sqrt 2)^{2/3}m_\sigma^{-1}$ of the $\sigma$-spin component into $\mu_\sigma$ in Eq. (\ref{HHH}), we find $h=0$. This is simply because the magnitude of the Fermi momentum in a free Fermi gas is independent of the particle mass at $T=0$. Actually, $h$ becomes finite at finite temperatures, because the temperature dependence of the chemical potential usually depends on $m_\sigma$. However, when we ignore all many-body fluctuation effects and simply treat the temperature effect within the mean-field theory, a finite $T_{\rm c}$ is obtained everywhere irrespective of the magnitude of $m_\uparrow/m_\downarrow$, as shown in Fig.\ref{fig4}(b). This is simply because the mismatch of the Fermi surfaces between the two components always disappears at $T=0$ within the mean-field level. Thus, the vanishing $T_{\rm c}$ seen in Fig.\ref{fig4}(a) is considered to come from an interaction effect beyond the mean-field level. Indeed, when the Fermi surface is smeared by a many-body effect, it is expected to give an effect similar to the temperature, leading to a finite $h$.
\par
We briefly note that, while the present ETMA involves higher order fluctuation effects than the ordinary $T$-matrix theory, it is still not fully self-consistent in the sense that the non-interacting Green's function $G^0$ is used in the pair correlation function in Eq. (\ref{PI}). In addition, we have assumed the second order phase transition, as well as the simplest uniform $s$-wave pairing state, in determining $T_{\rm c}$. Thus, since the vanishing $T_{\rm c}$, as well as the first order phase transition like behavior of $T_{\rm c}$ seen in Fig.\ref{fig4}(b), might depend on details of the theory, further analyses on these would be necessary, which remains as our future problem.
\par
\section{Summary}
To summarize, we have investigated the superfluid phase transition of a cold Fermi gas with mass imbalance. We have extended the $T$-matrix theory to include higher order fluctuation effects, so as to overcome the serious problems existing in the NSR theory and the $T$-matrix theory. Using this extended $T$-matrix theory, we have calculated $T_{\rm c}$ in the BCS-BEC crossover region to examine effects of mass imbalance. We show that, within ETMA, while a finite $T_{\rm c}$ is always obtained in the BEC regime, $T_{\rm c}$ vanishes at a certain value of $m_\uparrow/m_\downarrow$ in the weak-coupling BCS regime. Since the Fermi superfluids/condensates have been discussed, not only in the field of cold Fermi gas physics, but also in semiconductor physics (exciton), as well as in high-energy physics (color superconductivity), our results would be useful for the study of these hetero-type Fermi superfluids.

\begin{acknowledgements}
We thank S. Watabe and Y. Endo for useful discussions. This work was partially supported by Institutional Program for Young Researcher Oversea Visits from the Japan Society for the Promotion of Science. Y. O. was supported by Grant-in-Aid for Scientific research from MEXT in Japan (22540412, 23104723, 23500056).
\end{acknowledgements}


\end{document}